\documentclass[11pt]{article}
\usepackage{fullpage}
\usepackage{color}
\usepackage{graphicx}
\usepackage{epsfig}
\usepackage{url}

\twocolumn

\title{\LARGE \bf
Virtual Breakpoints for x86/64
}

\author{Gregory Price$^{1}$
\thanks{*This work was supported by Raytheon CSI}
\thanks{$^{1}$Gregory Price is {\tt\small price.gr at husky.neu.edu}}%
}

\begin{document}

\maketitle
\thispagestyle{empty}
\pagestyle{empty}

\begin{abstract}

Efficient, reliable trapping of execution in a program at the desired location is a linchpin technique for dynamic malware analysis.  The progression of debuggers and malware is akin to a game of cat and mouse - each are constantly in a state of trying to thwart one another.  At the core of most efficient debuggers today is a combination of virtual machines and traditional binary modification breakpoints (int3).  In this paper, we present a design for Virtual Breakpoints --- a modification to the x86 MMU which brings breakpoint management into hardware alongside page tables.  In this paper we demonstrate the fundamental abstraction failures of current trapping methods, and design a new mechanism from the hardware up.  This design incorporates lessons learned from 50 years of virtualization and debugger design to deliver fast, reliable trapping without the pitfalls of traditional binary modification.

\end{abstract}

\section{Introduction}

In the modern age of malware analysis and ``fuzzing for dollars'', security researchers still seek a robust, transparent, efficient trapping system.  The field heavily leverages virtualization technologies \cite{spider, stealthy, stealthy2, ether, baremetal, hardperf, overshadow, antidebug}.  Unfortunately, these technologies inherit the failings of traditional trapping methods while adding the complexity of architecture-specific implementations.

Debuggers \cite{gdb} and other instrumentation tools \cite{ttanalyze}\cite{bruening}\cite{ether}\cite{bintrans} are becoming increasingly complex as malware deploys equally complex anti-analysis techniques.  Many techniques require special memory allocators, compilers, or complex systems to control the ``view'' of memory \cite{antidebug, spider}.  In some cases, the efficiency or robustness of these techniques rely on architecture-level quirks that are more of a happy accident, rather than intentional design.  For example, SPIDER breakpoints \cite{spider} rely on a caching quirk of Intel CPU's to remain efficient, and many dynamic binary instrumentation systems \cite{antidebug} are a mess of trampolines and runtime hooking.

Despite the growing complexity, all of these systems must trade between efficiency, reliability, and transparency \cite{bruening}. If a trapping mechanism introduces significant overhead, analysis becomes tedious.  Traps that can be bypassed via detection or eviction inherently unreliable.  Traps that modify a target program's memory can cause corruption.  These trade-offs create a game of whack-a-mole, which suggests we are simply treating symptoms, rather than addressing the disease.

Trapping solutions such as single-stepping, debug registers, and full system emulation have been tried, but all fail to meet the flexibility and efficiency requirements to reach wide-spread adoption.  Single stepping a large, complex program (like an OS) is extremely inefficient \cite{gdb}\cite{spider}.  Usage of debug registers can be detected\cite{spider}.  Finally, pure-emulation is simply no longer sufficient to run and debug modern operating systems.

Researchers have made significant headway in solving reliability and efficiency problems by leveraging virtualization\cite{bruening, spider, ttanalyze, baremetal}, but most still rely on some form of binary modification.  The difficulty with binary modification for debugging hostile programs is the lack of assurance the modifications are not a) detected, b) bypassed, or c) introducing undefined behavior.  This is a result of these trapping mechanisms being designed prior to the advent of modern security and reverse engineering requirements.

``Cutting edge'' x86/64 debugging systems such as SPIDER \cite{spider} may accomplish (to an extent) the goal of stealth and efficiency, but still fail to mitigate the corruption and reliability problem with binary modification breakpoints.  Binary modification systems carry a presumption of well-behaved programs and lack of user error.  For example, if a 5-byte instruction is located at memory address 0x10, and a trap instruction is inserted at memory address 0x12, the resulting behavior could be a range of unintended consequences (Illegal Instructions, Jumps to Random Memory, Memory Corruption, etc).  SPIDER cannot handle this modification or user-error case directly, handling the degenerate case by simply evicting the trap and falling back to emulation (trading reliability for efficiency). 

In this paper, we propose a memory management unit extension for virtual machine debugging that addresses each of these core issues.  We claim that the corruption issue is the critical piece of the puzzle that, if solved, will lead to robust, efficient debuggers at both the system and virtual machine level.  The current field of debuggers attempt to fix this fundamentally unsolvable problem by way of building ``a better mousetrap'', but the problem they are attempting to solve reduces to the halting problem.

Our proposed solution introduces a ``breakpoint buddy-frame'' and adds a ``breakpoint bit'' to page table entries.  When a byte on a page with this bit set is accessed, the breakpoint frame is referenced (in hardware) to determine if that address has a breakpoint set.  This breakpoint information is stored on a byte-per-byte basis with the data frame.

For each byte read from a breakpointed page, an 8-bit value is retrieved from the breakpoint buddy-frame and used to determine if an interrupt is generated.  This byte implements standard read/write/execute breakpoint settings, and generates a debug break prior to executing the target instruction if the conditions are met.  The remaining bits remain open for future development.   

By doing away with binary modification as the de facto standard of breakpointing, we gain reliability, transparency, and guaranteed correctness of target program execution --- all without sacrificing flexibility or efficiency.


\section{Background}

Debuggers and dynamic analysis tools traditionally implement breakpoints in one of three ways: single-stepping, debug registers, and binary modification \cite{gdb}. Each mechanism is not without their flaws, and much research \cite{ttanalyze, bruening, stealthy, stealthy2, ether, baremetal, bintrans} has been done on the topic of mitigating the veritable list of issues.

In a review of debugging technology published in 1990\cite{impdebug}, Vernon Paxson lays out a list of techniques for debugging software that is eerily familiar.  Despite almost 30 years of research since then, few if any new hardware-supported debugging capabilities have been developed by major hardware manufacturers.  Even the ``new'' ARMv8.5 Memory Tagging Extensions (MTE) are not truly new, with Paxson discussing fully tagged architectures existing as far back as 1982 (possibly earlier).

Despite modern tools trying to mitigate the flaws of dated trapping mechanisms, each subsequent system increases complexity, reduces efficiency, or narrows in applicability.  All still fall prey to anti-debugging techniques such as timing attacks, code integrity checks, and just-in-time compilation or code relocation.

\subsection{Traditional Trapping Methods}

Each established method of trapping exhibits their own unique failures.  Each battles with trying to achieve flexibility, efficiency, transparency, and reliability - but none seem to solve for all four.

Single-stepping approaches make executing large, complex programs unbearably slow.  Typically implemented via the use of the eflags/rflags register, a full context switch between debugger and debuggee is required on each instruction.  This can push execution time to be orders of magnitude longer than the original program \cite{spider}.  Even emulation approaches, which can be viewed as a form of binary interpreter, are simply too slow for general use.

Debug Registers are an efficient but finite resource that are not practical nor easily virtualized. On x86 there are only 4 debug registers (DR0-DR4), limiting a debugger to 4 total watch/break points.  Further, because this is a physical register limitation, the debuggee can typically detect whether these registers are being used \cite{spider}\cite{antidebug}.

Finally, while binary modification meets the requirement of efficient execution, it is easily detected by a debuggee which monitors the integrity of its own code \cite{spider, antidebug, stealthy, stealthy2}.  On x86, these traps are implemented by placing an ``int3'' (debug break) instruction at the given address.  To see the problem with this technique, imagine a debuggee that periodically hashes its entire read-only codebase.  It would be able to detect this change, and modify its behavior.

It is still not apparent whether these traditional breakpointing methods can meet the requirements of an ``optimal'' solution.  In-fact, we claim binary modification produces an abstraction that may guarantee it can never provide an ``optimal'' solution.

\subsection{``Stealthy'' VM Breakpoints}

The first virtual machine breakpointing systems utilized binary modification as a primary trapping mechanism.  This carried with it all the typical headaches, namely that it is easily detected and may cause corruption.  Much research has been done to hide these breakpoints, but only recently has an efficient solution emerged \cite{spider}.

When Intel and AMD released virtualized Page Table support (known as Extended or Nested Page Tables \cite{intel, amd}), the idea of transparent binary modification came to fruition with SPIDER \cite{spider}.  SPIDER introduced a binary modification mechanism that made use of extended page tables to split the ``read-write'' view of guest data from the ``execute'' view of data.  By doing so, a guest cannot view a trap set via binary modification, because the guest may only read from a sanitized view of memory.

Unfortunately, this transparent breakpointing system still fails to be sufficiently flexible and reliable.  First, it falls victim to the ``Critical Byte Problem'' which will be described in the next section.  Second, because the host must maintain consistency between data and execution views, the guest still has a mechanism (writing to its code pages) with which to for a trap eviction.

Moving forward, we will demonstrate that systems relying on binary modification cannot achieve perfect flexibility and reliability.  If we hope to accomplish truly transparent and efficient breakpointing, then we must design it from the hardware up.  


\section{Overview}

In this section we discuss the goals of an ``optimal'' trapping system.  Next, we will construct an abstraction of how present-day trapping systems are implemented and identify violations of our goals.  What we find is that current systems are attempting to solve an impossible problem.

\subsection{Goals}

To start, we borrow the requirements laid out by SPIDER\cite{spider}, with one modification made to the requirement of Flexibility.

\begin{enumerate}
\item Flexibility	: Ability to set a trap \emph{at any memory address.}
\item Efficiency	: Maintain high performance
\item Transparency	: The target program should not be able to detect breakpoints
\item Reliability	: The target program should not be able to bypass or tamper with breakpoints.
\end{enumerate}

When we modify flexibility to state \emph{any memory address} instead of \emph{any instruction}, we find that no binary modification solution can accomplish all four requirements.  Even the best system, given a degenerate case, falls back on alternate trapping methods which violate at least one requirement.  For example, when a debuggee overwrites instrumented code under SPIDER, binary modification traps must be evicted from the modified page to avoid introducing undefined behavior.  We have dubbed this issue the "Critical Byte Problem".

\subsection{The ``Critical Byte Problem''}

Binary modification breakpoints depend on the execution of an instruction to trigger a breakpoint interrupt \emph{(Figure 1)}.  On x86, this is accomplished by replacing the first byte of an instruction with an int3 instruction (binary 0xCC) \emph{(Figure 2)}.  This dependence on \emph{execution of debuggee code} to determine \emph{debugger behavior} is a form of implicit trust.  The debugger trusts the binary will behavior nicely, while the debuggee trusts that the modification will not introduce undefined behavior.

\begin{figure}[h!t]
\begin{center}
\includegraphics[width=0.35\textwidth]{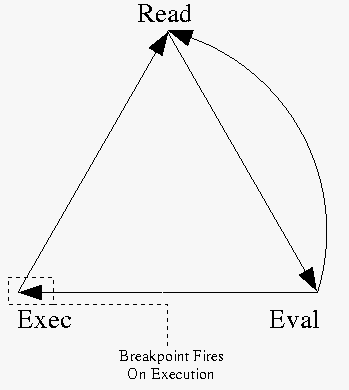}
\caption{\label{fig:virtlayer.png}Standard x86 processor Read-Eval-Execute loop.  Breakpoints fire on execute.}
\end{center}
\end{figure}

This causes what we have dubbed the ``Critical Byte Problem''.  If a binary modification trap is set on any byte other than the first of an instruction, it introduces undefined behavior.  \emph{Figure 2} shows an example of how a jmp instruction can be overwritten with an int3 instruction, or modified to jump to the incorrect place.

The processor reads and evaluates the first byte of an instruction to determine the general behavior and the full length of the instruction.   When a trap is aligned correctly, the instruction is effectively replaced in its entirety.  When a trap is misaligned, the 0xCC byte is decoded as part of the jump address.

\begin{figure}[h!t]
\begin{center}
\includegraphics[width=0.50\textwidth]{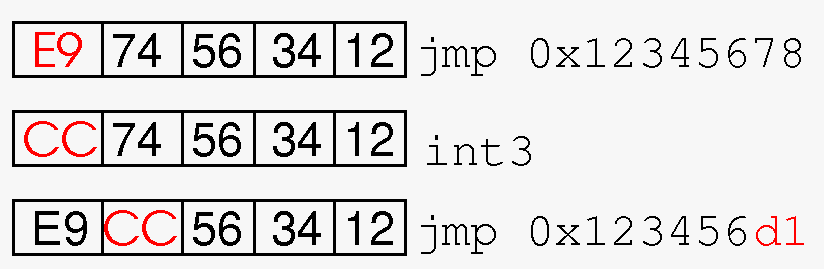}
\caption{\label{fig:criticalbyte.png}Misaligned breakpoints cause undefined behavior}
\end{center}
\end{figure}

What the \emph{figure 2} shows is that our ``optimal system'' should not depend on the execution of debuggee instructions, nor should it modify debuggee data.  We should avoid trusting the debuggee to provide debugger functionality, and we should maintain the integrity of debuggee data to guarantee correctness.

Systems such as SPIDER attempt to build complex solutions to this problem, but only end up trading efficiency for reliability.  Their system does not correct for user-error, and when instrumented code pages are modified, a ``Code Modification Handler'' is invoked --- usually to evict the trap and fall back to emulation.  Further, no binary modification breakpoint system can programmatically re-set a breakpoint over data that has been overwritten by the debuggee, as determining where in an arbitrary program is ``safe'' to instrument maps directly to the halting problem (the problem depends entirely on the behavior of the program).

What we find is that the critical byte problem is a symptom of a larger abstraction failure.  The next section will de-construct this failure, and show how providing a proper virtualization layer can mitigate the problem.

\subsection{Memory Abstractions}

It's useful to step back and look at how data in a virtualized system is segregated by ownership. First, in \emph{Figure 3}, we examine virtual memory implementation on x86.  The page tables, in software or in hardware, can be viewed as being owned by the Operating System.  Conversely, the contents of the data frame can be viewed as being owned by the software.  Most Operating Systems do not allow programs to directly modify page table contents for security reasons.  This concept of \emph{virtual memory} will be useful in designing our optimal breakpoint solution later.

\begin{figure}[h!t]
\centering
\includegraphics[width=0.40\textwidth]{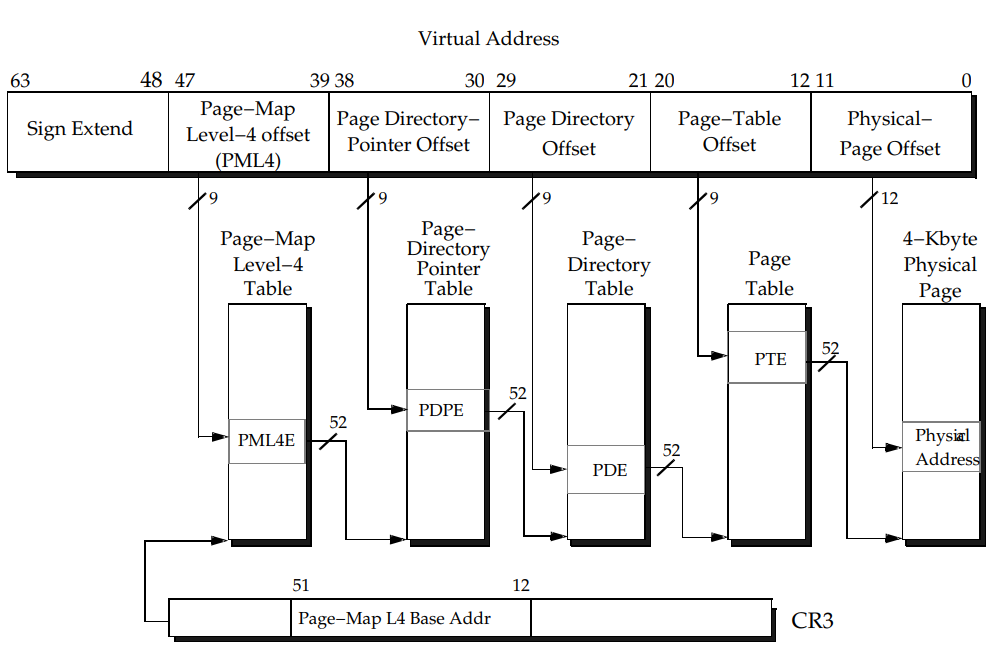}
\caption{\label{fig:virtual.png} Virtual Memory on x86}
\end{figure}

Virtualization extensions for x86 on both Intel \cite{intel} and AMD \cite{amd} processors provide a second layer of memory virtualization for virtual machines. Extended (Intel) or Nested (AMD) Page Tables, as in \emph{figure 4}, create two layers of page tables, where each guest memory reference must go through a second layer of translation to reach the physical memory frame.  These page tables are appropriately named the ``Guest Page Table'' and the ``Host Page Table''.

\begin{figure}[h!t]
\centering
\includegraphics[width=0.40\textwidth]{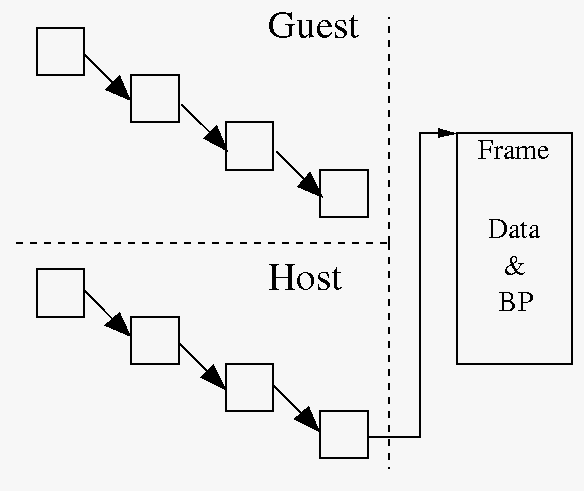}
\caption{\label{fig:currentview.png} Extended Page Tables still share a single frame for data and breakpoints}
\end{figure}

When we build a similar abstraction model for SPIDER (\emph{figure 5}), we see a pattern emerge when we focus on ``ownership'' of resources.

\begin{figure}[h!t]
\centering
\includegraphics[width=0.50\textwidth]{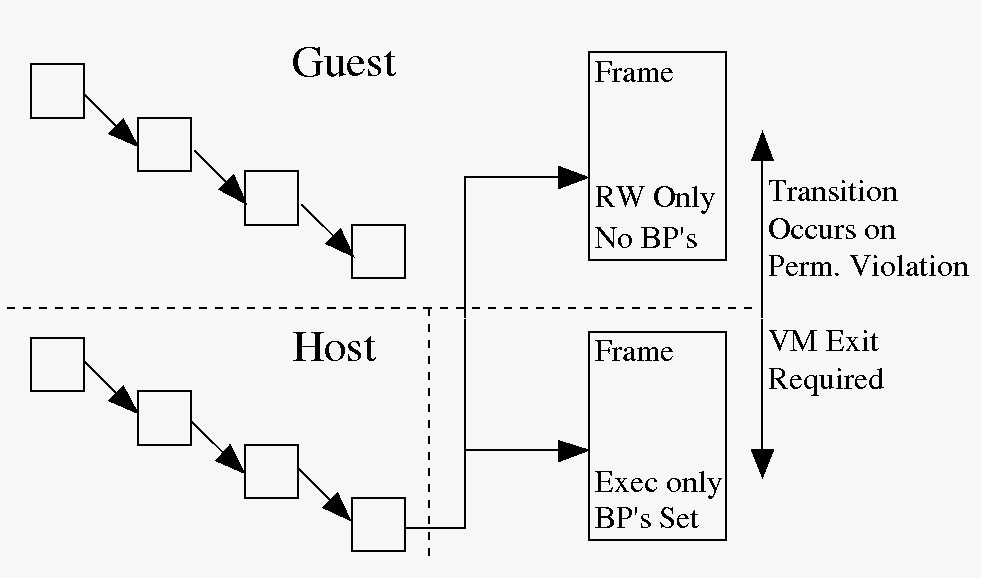}
\caption{\label{fig:spoderpoints.png} SPIDER also uses a frame containing data and breakpoints}
\end{figure}

In all three abstractions, we can see that debugger and debuggee data are contained within the same frame when binary modification traps are introduced.  A userland debugger, such as GDB, places int3 instructions directly into the program's data frame.  The same mechanism is used in naive virtual machine debugging, as seen in \emph{Figure 4}.  Likewise under SPIDER, despite the read/write view of program data being sanitized of debugger data, the execution view still holds both debugger and debuggee data.

Given this type of trap and debugger design design, an instrumented program can trivially read and/or interact with breakpoints --- or cause a change in debugger behavior (via eviction). A program debugged under SPIDER, despite not being able to see the breakpoints, can still force an eviction by blindly overwriting its own code pages (a simple code-migration technique may be sufficient).  There exists no mechanism with which to prevent this, as the program is trusted with complete control over its data.

In summary, even the most effective mechanism today fails to fully mitigate breakpoint eviction (either outright, or efficiently) due to placing debugger and debuggee data in the same data frame.  When a breakpoint overwrite occurs, current solutions must either fall back to another execution method, use a more complex trapping mechanism, or simply evict the breakpoint all together.

In the following section, we will propose a system which entirely separates debuggee data from debugger data (breakpoints).  This simplifies the handling of degenerate cases, mitigates the ``Critical Byte Problem'', and inherently provides transparency.


\section{Design}

When new hardware-supported trap mechanisms stopped appearing, malware analysis was not a global industry, and the challenges and requirements related to the work hardly dreamt of.  Given this, it is worth exploring extensions which may require new hardware support.  Our focus will be on solving the ``Critical-Byte Problem'' with a new virtualization extension to the x86 Memory Management Unit (MMU).

When modifying the memory management unit of a processor, some care must be taken to extend it in such a way that is reasonably friendly to existing operating systems.  Our design is no exception, so we must identify a solution that leverages existing Operating Systems structures.  Likewise, we must retain backwards compatibility with all existing trapping mechanisms.

\subsection{Breakpoint Virtualization Layer}

Unlike developing software breakpoint systems, a similar hardware-based breakpoint virtualization system has extremely strict limitations and requirements.  We claim any MMU modification that supports a separate view of data and breakpoints must adhere to (at a minimum) the following three requirements to have a chance at being adopted:  

\begin{enumerate}
\item TLB Compatibility
\item Hardware-Friendly Translation Mechanism
\item Flexible Allocation and Use
\end{enumerate}

To be TLB compatible, any design must be limited to using only the machine physical address and the page size.  When using a virtual machine, the use of host virtual addresses is prohibited because the TLB stores only a direct translation from guest virtual to machine physical address\cite{intel}.  We are afforded the page size from the MMU being set up by the operating system to walk a predetermined structure, however this could also be configurable via a new Model Specific Register.

Next, the mechanism of translation from a data address to a breakpoint address must be simple and fast.  Requiring multiple additional memory dereferences during an instruction or data fetch may not be feasible.

Finally, the mechanism must be optional.  Requiring constant use on all pages would effectively halve the size of usable memory and introduce significant overhead.  While this may be considered a fair trade off for a specialized system, a general purpose system must allow optional use.

\begin{figure}[h!t]
\centering
\includegraphics[width=0.50\textwidth]{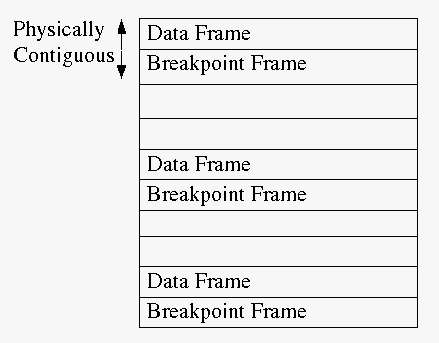}
\caption{\label{fig:buddyframe_phys.png} Individually contiguous buddy frames avoid the need for large contiguous areas of memory}
\end{figure}

When applying the first two requirements strictly, we find a design that uses physically contiguous page frames to be the obvious solution.  We call this a \emph{Buddy Frame}.  Implementing buddy frames on a per-frame basis, rather than in bulk, allows us to avoid allocating large contiguous areas memory (\emph{figure 6}).  This frame could be global (1 per physical frame) or per-cpu (1 buddy frame per physical or virtual cpu).

A per-cpu buddy frame system would require the use of the cpu id (0-n) during address translation to determine the offset from the data frame to buddy frame.  It may require additional TLB and Cache invalidation as well.  There are some complexities with a per-cpu option that are not considered within this work.  We leave this design extension up to an implementer to experiment with.  For the remainder of this work, we will work with a single buddy frame design.

\begin{figure}[h!t]
\centering
\includegraphics[width=0.50\textwidth]{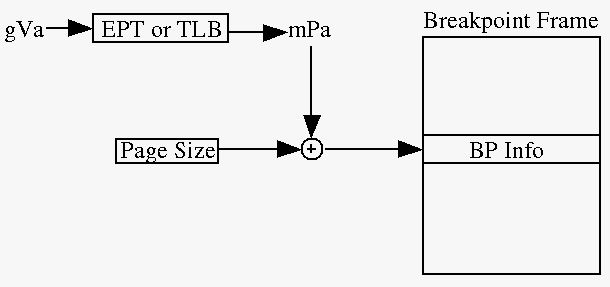}
\caption{\label{fig:VBP_Layer_BuddyFrames.png}Virtualization layer to translate physical address to buddy frame addresses}
\end{figure}

The translation from data address to breakpoint address is then a simple addition of the page-size (\emph{figure 7}).  This also makes our solution extensible to operating systems configured to use pages larger than the standard 4KB as long as this information is retrievable in hardware by the MMU.

The last requirement dictates an agreement between the operating system and the MMU about how page frames will be allocated and managed.  This is accomplished through the use of a page table entry which has software-defined bits.

\begin{figure}[h!t]
\centering
\includegraphics[width=0.40\textwidth]{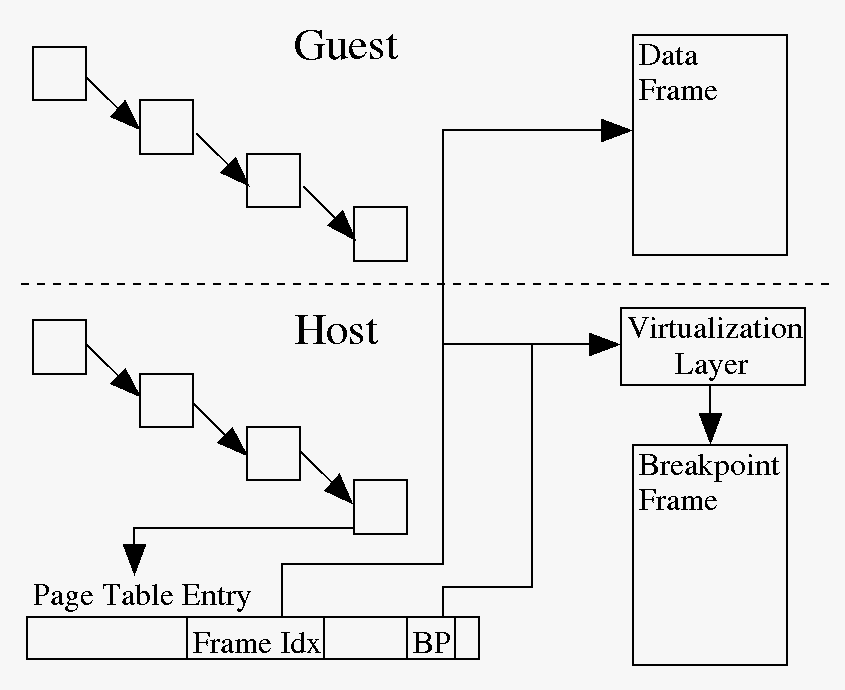}
\caption{\label{fig:virtbp.png}Virtual Breakpoints.  Data is ``debuggee-only'', Breakpoints are ``debugger-only''.}
\end{figure}

We propose the addition of a ``Breakpoint Bit'' to the page table entry \emph{(figure 8)}.  Much like the ``present bit'' determines whether the MMU produces a page-fault, the breakpoint-bit would determine whether the MMU would do a subsequent breakpoint-frame lookup for an accessed address.  Breakpoint info would be stored in byte-for-byte parity with the data frame.  \emph{Figure 8} also shows that our new ``ownership'' abstraction has resolved the issue of debugger and debuggee data living within the same data frame.

This design can be seen as applying the idea of Virtual Memory \emph{(Figure 3)} to the concept of breakpoints.  This is the provenance of the name \emph{Virtual Breakpoint}.

\subsection{OS Modifications}

Given the above hardware implementation, the OS modifications to support virtual breakpoints are straight-forward.  An Operating System must provide four new mechanisms:

\begin{enumerate}
\item Buddy Frame Allocation
\item Page Table Entry ``Breakpoint Bit''
\item Breakpoint Manipulation API
\item Buddy Frame Swap Compatibility
\end{enumerate}

The first two mechanisms are solved by implementing a new kernel allocator option.  This option would allocate two physically contiguous frames, and set the breakpoint bit on the data frame's page table entry.  Buddy Frame Allocation may be flexibly applied if Copy-On-Write is required for an instrumented page.  The inheriting task may either automatically allocate its own buddy page during Copy-On-Write, or it may leave it behind.  The author notes that this is an example of how such a hardware extension may open new opportunities for debugger design.

The third mechanism is a new kernel API which allows for the breakpoint frame to be accessed in a pre-defined manner by the debugger.  A per-byte breakpoint API could be used for precision, and a per-page optimization could be provided if large breakpoints are needed.  This API gives debugger and instrumentation developers a standardized way to set breakpoints.

This system places no limitation set on accessibility of debuggee data by the debugger, traditional binary modification breakpoints can still be used.  The retains backward compatibility for all existing debugging platforms.

Finally, in a naive system, care must be taken not to swap out buddy frames when their data-frame counterparts are still in memory (and vice-versa).  The operating system's swapping algorithm could account for this by pinning the frame or by utilizing an additional bit in the Page Table Entry that denotes whether a given frame is itself a breakpoint frame.  This work leaves this as future work for OS and debugger designers.

\subsection{Analysis of Design}

Using the above design, we re-examine the original design goals we set out to accomplish.

\begin{itemize}
\item Flexibility: Breakpoints can be set on any address, and the allocation of buddy-frames is limited to just those pages where breakpoints are set.

At worst the number of frames can increases linearly with the number of breakpoints.  This can double memory usage.  However, we note that SPIDER breakpoints suffer from the same problem \cite{spider}.

\item Efficiency: Debuggee execution now occurs purely in bare-metal without requiring additional instrumentation for misbehaved programs.

A debuggee now executes and performs I/O on the same data frame without incurring a context switch.  A context switch on systems such as SPIDER can take thousands of instructions \cite{spider}, and occurs twice per fault (guest exit and entry).  Instead, it is possible this solution may reduce this cost by distributing a small number of cycles per instruction on an instrumented page.

When the number of breakpoints is small and concentrated to a small portion of code (a typical use-case), performance suffers only when utilizing an instrumented page.  In fact, performance should be no worse than SPIDER's stealthy breakpoints for self-modifying code given the similar design. As the number of breakpoint frames grow, the number of memory references grows by a factor of at most 2 (one for data, one for breakpoint information).

\item Transparency:  Complete segregation of debuggee and debugger data ensures the debuggee has no avenue with which to see breakpoint information.  

\item Reliability:  Complete segregation of debuggee and debugger data ensures the debuggee has no avenue with which to affect breakpoint information.
\end{itemize}

In the worst case scenario, a breakpoint is set on every page of a target.  This doubles memory usage and causes an additional memory reference to occur per memory reference.  In both cases, existing systems exhibit at least as bad performance, if not worse.  SPIDER breakpoints double memory usage per utilized page and can cause thousands of instructions of overhead per memory access on an instrumented page.


\section{Implementation}  

This section provides a roadmap for implementing a prototype of a virtual breakpoint system.  At first, we propose implementing the solution in an emulator to prove feasibility, and then extending an existing hardware platform (potentially via FPGA) to prove efficiency.  However, notably, neither implementation would be useful beyond further academic exploration of the problem, since extending an emulator would lead to severe inefficiencies and existing open-architectures tend not to exhibit the critical byte problem.

\subsection{Modify QEMU i386 MMU}

\begin{itemize}
\item tlb pte breakpoint bit checking
\item page table walking bit checking
\item buddy frame lookups when breakpoint bit is set
\item raise an exception when breakpoint terms are met
\end{itemize}

\subsection{Modified Linux Kernel}

\begin{itemize}
\item page table entry breakpoint-bit modification
\item breakpoint page allocation option/interface for buddy frames
\item breakpoint management interface (set breakpoint on given address)
\item virtual breakpoint interrupt handler
\item add memory mapping modifier (modify non-breakpointed page to add virtual breakpoint page)
\item memory swap algorithm modification to avoid swapping buddy-pages
\end{itemize}

\subsection{Modified Debugger}
\begin{itemize}
\item add/remove virtual breakpoint page command
\item add/remove virtual breakpoint command
\item add/remove virtual watchpoint command
\item virtual breakpoint signal handler
\end{itemize}


\section{Future Work}

By solving the critical-byte problem, we open up a number of possibilities - including new types of breakpoints, and implementing more efficient code coverage systems.  This section discusses this potential extensions.

\subsection{New breakpoint types}
\begin{itemize}
\item Break on Instruction Fetch

Presently, instruction-fetch breakpoints are not possible (except for work-around like debug registers and page table permission twiddling).  A bit in the 8-bit breakpoint field on a buddy-page could be utilized to cause a breakpoint interrupt to fire on instruction fetch.

\item Nesting 

Implementing the virtualization layer such that a guest can set up another layer of virtual breakpointing.  The would allow for introspection of a hostile virtual machine within another virtual machine.

\item Coverage/Taint

A taint-propagation debugger can be developed that copies breakpoint-bits whenever data is transferred from one area of memory to another.  For example, When external data is injected into memory (via inbound network traffic), the storage location has coverage bits set in its buddy frame.  As this data propagates to other areas of memory, the coverage bits propagate with them. 

\item "Hook Points" 

A specialized interrupt handler could be registered with the operating system that is called whenever a ``hook point'' interrupt is fired.  The ``Hook Point'' is set on an address, and the address is entered into a map accessed by the interrupt handler.  Upon executing / accessing the hooked address, the debuggee exits directly to the interrupt handler for dispatching.  This is exceptionally useful for virtual machines, where dispatching could be handled directly upon VMExit (similar to a vmcall).

This could be useful for recording inbound or outbound non-deterministic events for checkpoint/restart and replay systems (such as clock reads via RDTSC).

\end{itemize}


\section{Related Work}

The idea of segregating ``views'' of data is not necessarily novel.  A number of technologies\cite{spider} \cite{stealthy} \cite{stealthy2} \cite{overshadow} employ software to modify shadow page tables, extended page tables, and interrupt handlers to mask certain data from guest machines. 

Overshadow\cite{overshadow} provides a mechanism with which to hide (encrypt) the contents of a guest-process's memory from the guest kernel.  The researchers goals were to provide a mechanism with which a guest process could execute securely, even if the guest operating system was actively hostile.  Overshadow accomplished this by extending the original form of page table virtualization released by Intel and AMD - Shadow page tables.  

Overshadow implemented a shadow page table containing multiple mappings (encrypted and unencrypted) of a guest's physical memory, and actively tracks the ``identity'' of the guest process attempting to read a page.  When the accessing process does not have the correct identity, an encrypted page is presented (on read) or the machine is terminated (on write).  When the accessing process does have the correct identity, an unencrypted page is presented with the permissions originally granted to the page.  Unfortunately, its dependence on shadow page tables means it's likely to be too inefficient \cite{vmware} for modern operating systems and heavy load systems.

VAMPiRE\cite{stealthy} takes advantage of virtual memory page permissions to be executed via alternate methods.  In particular, the developers chose to implement a single-step handler which executes any instruction (or accesses any data) falling on a page which contains a breakpoint.  This is done to ensure any reads and overwrites of breakpoints are not possible, and so the integrity of guest data is preserved.  Unfortunately, relying on single-stepping and what are effectively page-length breakpoints limits the flexibility of the breakpoint system.  For example, if a program is contained entirely within a single page, or a breakpointed page contains ``hot'' code, then speed will suffer dramatically.

Finally, Spider\cite{spider} comes the closest to implementing a solution that maximizes efficiency, while retaining the flexibility of traditional binary modification breakpoints.  Similar to VAMPiRE, it leverages virtual page permissions to determine what ``view'' of memory should be provided to the debuggee.  On read/write, a ``sanitized'' view of memory (sans breakpoints) is provided to hardware to prevent detection.  On execute, the modified page (containing int3 instructions) is provided for execution.  

Spider maintains efficiency by leveraging a quirk in caching mechanisms, which allows both views to be cached (one in the instruction cache, one in the data cache), limiting the number of VM Exits required to expose the correct data on a given access.  Unfortunately, as previously discussed, Spider still falls prey to some forms of classic anti-debugging techniques (such as overwriting) due to its reliance on binary modification.  

Each of these works provide a trend of researchers attempting to split the view of data based on whether the debuggee is trusted.  In Overshadow, some parts of the debuggee are trusted but not others.  In VAMPiRE and Spider, the debuggee is not trusted to read or write instrumented memory.  Recognizing the trust violation was the primary inspiration for producing fully segregated data and breakpoint frames.


\section{Conclusion}

In this paper we demonstrated the problems with traditional and modern introspection systems to show a clear need for a fully virtualized breakpoint solution.  As we broke down various breakpointing techniques, we identified two fundamental issues with current solutions.  Either these solutions were inefficient, or they traded integrity and correctness of execution for efficiency.  

We proposed a virtualization extension for the memory management unit.  We build this from the ground up with the goals of an optimal system in mind.  Our solution is novel in that it treats debuggee execution as an fundamentally untrusted action, and segregates all debug related information into a separate buddy-frame accessible only by the debugger.  Unlike other solutions which make heavy use of binary modification, our system retains integrity of breakpoints and correctness of execution by not requiring the modification of debuggee data.

We believe that this Virtual Breakpointing design the solution that virtual machine introspection and interposition systems have been looking for.  It presents opportunities for the development of brand new types of breakpoints, and is efficient enough to run commodity operating systems under test.



\begin{thebibliography}{99}

\bibitem{gdb}  Gdb. http://www.gnu.org/software/gdb/.

\bibitem{spider} Deng, Z., Zhang, X., \& Xu, D. (2013, December). Spider: Stealthy binary program instrumentation and debugging via hardware virtualization. In Proceedings of the 29th Annual Computer Security Applications Conference (pp. 289-298). ACM.

\bibitem{ttanalyze} U. Bayer, C. Kruegel, and E. Kirda. Ttanalyze: A tool for analyzing malware. In EICAR’06

\bibitem{bruening} D. Bruening. Efficient, transparent, and comprehensive runtime code manipulation. PhD thesis, 2004.

\bibitem{stealthy} Vasudevan, A., \& Yerraballi, R. (2005, December). Stealth breakpoints. In Computer security applications conference, 21st Annual (pp. 10-pp). IEEE.

\bibitem{stealthy2} Vasudevan, A. (2009, October). Re-inforced stealth breakpoints. In Risks and Security of Internet and Systems (CRiSIS), 2009 Fourth International Conference on (pp. 59-66). IEEE.

\bibitem{ether} Dinaburg, A., Royal, P., Sharif, M., \& Lee, W. (2008, October). Ether: malware analysis via hardware virtualization extensions. In Proceedings of the 15th ACM conference on Computer and communications security (pp. 51-62). ACM.

\bibitem{baremetal} Willems, C., Hund, R., Fobian, A., Felsch, D., Holz, T., \& Vasudevan, A. (2012, December). Down to the bare metal: Using processor features for binary analysis. In Proceedings of the 28th Annual Computer Security Applications Conference (pp. 189-198). ACM.

\bibitem{hardperf} Vogl, S., \& Eckert, C. (2012, April). Using hardware performance events for instruction-level monitoring on the x86 architecture. In Proceedings of the 2012 European Workshop on System Security EuroSec (Vol. 12).

\bibitem{antidebug} R. R. Branco, G. N. Barbosa, and P. D. Neto. Scientific but not academical overview of malware anti-debugging, anti-disassembly and anti-vm technologies. Blackhat USA’12.

\bibitem{bintrans}  P. Feiner, A. D. Brown, and A. Goel. Comprehensive kernel instrumentation via dynamic binary translation. In ASPLOS’12

\bibitem{intel} Intel, Intel. "Intel(R) 64 and IA-32 architectures software developer’s manual." Volume 3A: System Programming Guide, Part 1.64 (64).

\bibitem{amd} Amd, Amd.  "Developer Guides, Manuals, \& ISA Documents." https://developer.amd.com/resources/developer-guides-manuals/.

\bibitem{overshadow} Chen, X., Garfinkel, T., Lewis, E. C., Subrahmanyam, P., Waldspurger, C. A., Boneh, D., ... \& Ports, D. R. (2008, March). Overshadow: a virtualization-based approach to retrofitting protection in commodity operating systems. In ACM SIGARCH Computer Architecture News (Vol. 36, No. 1, pp. 2-13). ACM.

\bibitem{vmware} Adams, K., \& Agesen, O. (2006). A comparison of software and hardware techniques for x86 virtualization. ACM SIGOPS Operating Systems Review, 40(5), 2-13.

\bibitem{impdebug} Paxson, V. (1990). A Survey of Support for Implementing Debuggers.

\bibitem{jeffk} Jeffk.

\bibitem{albert} Albert.

\end{thebibliography}
\end{document}